\documentclass{JHEP3}
\usepackage{amsmath}

\def\({\left(}
\def\){\right)}
\def\[{\left[}
\def\]{\right]}
\def\<{\langle}
\def\>{\rangle}
\def\lv{\left|}
\def\rv{\right|}
\def\und{\underline}
\def\ov{\overline}
\def\e{\varepsilon}
\def\de{\delta}
\def\la{\lambda}
\def\tla{\tilde{\lambda}}
\def\teta{\tilde{\eta}}
\def\ie{{\it i.e.,} }
\def\s{\hat{s}\,}

\newcommand{\be}{\begin{equation}}
\newcommand{\ee}{\end{equation}}
\newcommand{\barr}{\begin{array}}
\newcommand{\earr}{\end{array}}
\newcommand{\bal}{\begin{aligned}}
\newcommand{\eal}{\end{aligned}}

\newcommand{\labell}[1]{\label{#1}}

\title{Soft Theorem of $\mathcal{N}=4$ SYM in Grassmannian Formulation}
\author{Junjie Rao$^a$\footnote{Email: raojunjie@zju.edu.cn}\\
{$^a$Zhejiang Institute of Modern Physics, Zhejiang University, Hangzhou, 310027, P. R. China}}

\abstract{Inspired by the new soft theorem in gravity by
Cachazo and Strominger, the soft theorem for color-ordered
Yang-Mills amplitudes has also been identified by Casali.
In this note, the same content of $\mathcal{N}=4$ SYM using the Grassmannian formulation is studied.
Explicitly, in the holomorphic soft limit,
we reduce the $n$-particle amplitude in terms of Grassmannian contour integrations
into the deformed $(n-1)$-particle amplitude by localizing $k$ variables relevant to the $n$-th particle.
Afterwards, the leading soft factor and sub-leading soft operator naturally emerge.}


\begin{document}
\maketitle

\newpage
\section{Introduction}
\labell{intro}

Scattering amplitudes often have an universal soft behavior when the momentum of one
external leg tends to zero. This soft limit can be traced back to the works \cite{Low:54,Low:58,Weinberg}.
Recently, a new soft theorem for tree level gravity amplitudes was studied in \cite{Cachazo:14soft}.
By using the BCFW construction and imposing the holomorphic soft limit,
Cachazo and Strominger have proved that
\be
\bal
&M_n\(\la_n\to\e\la_n\)\\
=&\frac{1}{\e^3}\sum_{a=1}^{n-2}\frac{\<n-1,a\>^2[na]}{\<n-1,n\>^2\<na\>}
M_{n-1}\(\tla_{n-1}\to\tla_{n-1}+\e\frac{\<an\>}{\<a,n-1\>}\tla_n,\,\tla_1\to\tla_1
+\e\frac{\<n-1,n\>}{\<n-1,a\>}\tla_n\)+O(\e^0),
\eal
\ee
here for $M_n$ and $M_{n-1}$, the unmentioned external kinematic data are un-deformed
and we prefer to suppress them for conciseness\footnote{Any amplitude in this note
contains the delta function of momentum conservation.}.
Taylor expansion in $\e$ exhibits three singular terms in orders $\e^{-3}$, $\e^{-2}$ and $\e^{-1}$,
while higher order terms in $\e$ will be mixed with the less interesting $O(\e^0)$ parts.

A similar relation for tree level Yang-Mills amplitudes using the BCFW construction,
proved by Casali \cite{Casali:14soft}, takes the form
\be
\bal
&A_n\(\la_n\to\e\la_n\)\\
=&\frac{1}{\e^2}\frac{\<n-1,1\>}{\<n-1,n\>\<n1\>}
A_{n-1}\(\tla_{n-1}\to\tla_{n-1}+\e\frac{\<1n\>}{\<1,n-1\>}\tla_n,\,\tla_1\to\tla_1
+\e\frac{\<n-1,n\>}{\<n-1,1\>}\tla_n\)+O(\e^0), \labell{eq-16}
\eal
\ee
where two singular terms in orders $\e^{-2}$ and $\e^{-1}$ appear
after Taylor expansion. The mixing between higher order terms from the
deformed $A_{n-1}$ and $O(\e^0)$ parts also pertains to this case.

It is fruitful to study the same object from various viewpoints in physics because it will
deepen our original understanding and reveal many hidden facts.
Many related studies have been achieved including:
the soft limit from Poincar\'{e} symmetry and gauge
invariance \cite{BLPR:14constrain,BDVN:14lowenergy},
Feynman diagrams approach \cite{White:14diagram},
conformal symmetry approach in Yang-Mills theory \cite{Larkoski:14conformal},
the soft limit in arbitrary dimensions
\cite{SV:14arbitrary,Afkhami-Jeddi:14arbitrary,Zlotnikov:14arbitrary,KR:14arbitrary},
loop corrections of the soft limit \cite{BDN:14oneloop,HHW:14loop,Cachazo:14renormalized,BHHW:14more},
string theory approach \cite{Schwab:14string,BHHW:14more},
ambitwistor string approach \cite{ACS:14null,GLM:14null} and KLT approach \cite{DFFW:14KLT}.

In this note, a relatively novel way using the Grassmannian contour integral
is shown to reproduce the soft theorem for amplitudes in $\mathcal{N}=4$ SYM.
Relevant background on Grassmannian can be found in
\cite{Arkani-Hamed:09duality,Arkani-Hamed:09unification,NVW:09etude,BTVW:10trees} and
a brief review of key formulae related to N$^{k-2}$MHV amplitudes $A^{[k]}_n$ is given here.
The first example is the NMHV (super)amplitude, written as
\be
\bal
&A^{[3]}_n=\int_{\{f_6=\ldots=f_n=0\}}\frac{g^{[3]}_n}{(n-1)(1)(3)}\frac{1}{f_6\ldots f_n},\\
&g^{[3]}_n=\prod_{j=6}^{n-1}(1\,2\,j)(2\,3\,j-1),~f_l=(l-2\,l-1\,l)(l\,1\,2)(2\,3\,l-2), \labell{eq-14}
\eal
\ee
where $(l)$ is short for the consecutive 3-minor $(l\,l+1\,l+2)$ in terms of $c_{Ii}$'s for the $k=3$ case.
The integral symbol above denotes
\be
\int d^{3(n-3)}c_{Ii}\,\de^{2(n-3)}(\la_i-\la_Ic_{Ii})\de^{2\cdot3}(\tla_I+c_{Ii}\tla_i)
\de^{4\cdot3}(\teta_I+c_{Ii}\teta_i),
\ee
note that the supersymmetric content is not involved in solving $c_{Ii}$'s.
There are $(n-5)$ actual integration variables to be localized\footnote{For general N$^{k-2}$MHV amplitudes,
the $(k\times n)$ matrix $c_{Ii}$ has GL$(k)$ gauge invariance, hence there are $k(n-k)$
independent variables. Imposing $(2n-4)$ delta functions, there are
$(k-2)(n-k-2)$ variables left to be fixed by contour integrations.}
by $(n-5)$ $f_i$'s in \eqref{eq-14}. Although the integrand in \eqref{eq-14} is nothing but
\be
\frac{1}{(1\,2\,3)(2\,3\,4)\ldots(n\,1\,2)}, \labell{eq-15}
\ee
after the cancelation of non-consecutive minors between the numerator and denominator,
for practical calculations \eqref{eq-14} is adopted and the reason is:
It is known that residue (or contour) integrations demand non-consecutive minors for
physical outputs, while some consecutive minors are redundant.
The advantage of the integrand in \eqref{eq-14} is that all unwanted
minors are `lifted away' from the sequence of minors mapped to zero
(\ie they are only evaluated as residues at zero minors).

Another convenience of this integrand is that in process of the inverse
soft operation (or `add one particle at a time'), the soft factor is
recovered in the soft limit. More explicitly, this factor in terms of minors is given
via $I^{[3]}_n=I^{[3]}_{n-1}S^{[3]}_{(n-1)\to n}$ where $I^{[3]}_n$
is short for the integrand in \eqref{eq-14}, so that
\be
S^{[3]}_{(n-1)\to n}=\frac{(n-2)'(n-1)'(n-2\,2\,3)}{(n-1)f_n}
=\frac{(n-2)'(n-1)'}{(n-2)(n-1)(n)},
\ee
here the prime means that the corresponding consecutive minor is for the $(n-1)$ case, namely
$(n-2)'=(n-2\,n-1\,1)$ and $(n-1)'=(n-1\,1\,2)$. In the limit $\la_n\to\e\la_n$,
three $c_{In}$'s are localized by two delta functions and one contour integration, which leads to
\be
S^{[3]}_{(n-1)\to n}\to\frac{1}{\e^2}\frac{\<n-1,1\>}{\<n-1,n\>\<n1\>},
\ee
this is the desired soft factor. In
\cite{Arkani-Hamed:09unification} the leading singular term is
mentioned, while in fact the sub-leading term is also automatically
included, as will be demonstrated in this note.

Having explained the integrand for NMHV amplitudes, now let's present
the universal structure of general N$^{k-2}$MHV
amplitudes derived by the conjugation construction. Before this, one can compare the integrand for
NMHV amplitudes \eqref{eq-14}, with the one for N$^2$MHV amplitudes given by
\be
A^{[4]}_n=\int_{\{F_7=\ldots=F_n=0\}}\frac{g^{[4]}_n}{(n-1)(1)(3)}\frac{1}{F_7\ldots F_n}, \labell{eq-21}
\ee
where
\be
g^{[4]}_n=\prod_{j=7}^{n-1}(1\,2\,3\,j)(2\,3\,j-2\,j-1)(1\,j-2\,j-1\,j)
\prod_{j=4}^{n-3}(1\,3\,j\,j+1)(1\,2\,j\,j+3),
\ee
and $F_l=f_{l1}f_{l2}$ with
\be
\bal
&f_{l1}=(l-3\,l-2\,l-1\,l)(l-3\,l\,1\,2)(l-3\,2\,3\,l-2),\\
&f_{l2}=(1\,l-2\,l-1\,l)(1\,l\,2\,3)(1\,3\,l-3\,l-2). \labell{eq-22}
\eal
\ee
The conjugation construction is implied in process of getting $g_n^{[4]}$ and $F_l$'s
from $g_n^{[3]}$ and $f_l$'s, as well as transforming $(n-1)(1)(3)$ of $k=3$ into the same product but of $k=4$.
By extending this construction, one can show that for general N$^{k-2}$MHV amplitudes,
\be
A^{[k]}_n=\int_{\{F_{k+3}=\ldots=F_n=0\}}\frac{g^{[k]}_n}{(n-1)(1)(3)}
\frac{1}{F_{k+3}\ldots F_n},~F_i=f_{i1}\ldots f_{i,k-2},
\ee
here each $F_i$ is a product of $(k-2)$ $f_{ij}$'s which enforce
$(k-2)$ minors to be zero, hence offsetting the $(k-2)$ variables brought by each newly added
particle. Its details can be found in appendix \ref{app1}.

Returning to the inverse soft operation, we need to highlight a relation. Assume that
the integrand $I^{[k]}_{n-1}$ is known, to get $I^{[k]}_n$
one simply needs to multiply $I^{[k]}_{n-1}$ by the inverse soft factor
\be
S^{[k]}_{(n-1)\to n}=\frac{(n-k+1)'(n-k+2)'\ldots(n-1)'}{(n-k+1)(n-k+2)\ldots(n-1)(n)}, \labell{eq-19}
\ee
which has been `over-simplified' due to the cancelation of all
non-consecutive minors between the numerator and denominator, same
as what happens in \eqref{eq-15}, which is the simplicity
in handling the soft limit: We need only focus on the consecutive minors.

After getting familiar with the Grassmannian formulation, we will
show how to use it to reproduce the supersymmetric extension of \eqref{eq-16} for
all $k$'s, especially two key components: the overall
soft factor, and the deformed anti-holomorphic spinor pair $(\tla_{n-1},\tla_1)$
along with its deformed supersymmetric counterpart $(\teta_{n-1},\teta_1)$,
which nicely imitates the former.

This note is organized as follows.
Section \ref{sec2} reviews the inverse soft operation for NMHV amplitudes
and proves the corresponding soft theorem. Section \ref{sec3} explores
the same aspects of N$^2$MHV and N$^3$MHV amplitudes and attempts
to find the pattern for general $k$'s. Section \ref{sec4} provides a general
proof of the soft theorem for N$^{k-2}$MHV amplitudes. Section \ref{conclu}
concludes with a few comments.
Appendix \ref{app1} derives the general structure for N$^{k-2}$MHV amplitudes
by using the conjugation construction. Appendix \ref{app2} explains
why the unmentioned but possibly singular parts are actually regular in the soft limit.

\section{NMHV Amplitudes Redux}
\labell{sec2}

In this section let's consider the simplest case, \ie the NMHV amplitude formulated by
\be
A^{[3]}_n=\int\frac{g^{[3]}_n}{(n-1)(1)(3)}\frac{1}{\und{f_6}\ldots\und{f_{n-1}f_n}}, \labell{eq-6}
\ee
where to simplify the notation, underlines are used to indicate
the zero factors for contour integrations.
The global residue theorem manipulates above expression into
\be
\bal
&\int\frac{g^{[3]}_n}{(n-1)(1)(3)}\frac{1}{\und{f_6}\ldots\und{f_{n-1}f_n}}\\
=&-\int\frac{g^{[3]}_n}{\und{(n-1)}(1)(3)}\frac{1}{\und{f_6}\ldots\und{f_{n-1}}f_n}
-\int\frac{g^{[3]}_n}{(n-1)\und{(1)}(3)}\frac{1}{\und{f_6}\ldots\und{f_{n-1}}f_n}
-\int\frac{g^{[3]}_n}{(n-1)(1)\und{(3)}}\frac{1}{\und{f_6}\ldots\und{f_{n-1}}f_n}, \labell{eq-2}
\eal
\ee
as will be explained in appendix \ref{app2}, among three terms above,
only the $\und{(n-1)}$ term has singular contribution in the soft limit $\la_n\to\e\la_n$.

To work out the integration, let's write the integral symbol explicitly as
\be
\int d^{3(n-3)}c_{Ii}\,\de^{2(n-3)}(\la_i-\la_Ic_{Ii})\de^{2\cdot3}(\tla_I+c_{Ii}\tla_i)
\de^{4\cdot3}(\teta_I+c_{Ii}\teta_i). \labell{eq-1}
\ee
In the Grassmannian formulation of $c_{Ii}$'s, one needs to
choose a gauge. Among many choices, the following gauge provides maximal simplicity:
\be
C=\(\barr{ccccccc}
\ldots & 1 & 0 & c_{n-2,n} & 0 & c_{n-2,2} & \ldots\\
\ldots & 0 & 1 & c_{n-1,n} & 0 & c_{n-1,2} & \ldots\\
\ldots & 0 & 0 & c_{1n} & 1 & c_{12} & \ldots
\earr\),
\ee
where three columns $(n-2,n-1,1)$ have been fixed to be a unit
matrix\footnote{The gauge choice is a bit different from the standard one
where columns of negative helicities are often fixed, since under such a choice the
supersymmetric counterpart can be integrated over most conveniently \cite{Arkani-Hamed:09duality}.
Here the three columns chosen to be fixed are not necessarily associated with negative helicities.
But for the soft particle $n$, which has positive helicity,
it is natural to have an unfixed column.}.

Now we attempt to write the integrand possessing $\und{(n-1)}$
in \eqref{eq-2} into a form of the deformed $A^{[3]}_{n-1}$ multiplied
by the soft factor, hence the integral can be split as
\be
-\int d^{3(n-4)}c_{I\und{i}}\,\de^{2(n-4)}(\la_{\und{i}}-\la_Ic_{I\und{i}})
\de^{2\cdot3}(\tla_I+c_{I\und{i}}\tla_{\und{i}}+c_{In}\tla_n)I^{[3]}_{n-1}
\left\{\int d^3c_{In}\de^2(\la_n-\la_Ic_{In})
\frac{(n-2)'(n-1)'}{(n-2)\und{(n-1)}(n)}\right\}, \labell{eq-3}
\ee
where $\und{i}=1,\ldots,n-1$, and
integrations over all $c_{In}$'s are collected inside the curly bracket.
Note that the supersymmetric content has been set aside temporally since it does not
provide localization constraints.

Keep in mind that the integrand inside the
curly bracket is the inverse soft factor $S^3_{(n-1)\to n}$,
which later turns into the soft factor when the residue is evaluated at $(n-1)=0$.
To reveal this, let's compute the relevant minors as
\be
(n-2)'=1,~(n-1)'=c_{n-2,2},
\ee
\be
(n-2)=c_{1n},~(n-1)=-c_{n-2,n},~
(n)=-\lv\barr{cc}
c_{n-2,n} & c_{n-2,2}\\
c_{n-1,n} & c_{n-1,2}
\earr\rv,
\ee
and the integration measure is defined to be (be aware of the reversed
cyclic order\footnote{The reason to choose this order will be explained in section \ref{sec4}.})
\be
\int d^3c_{In}\equiv\int dc_{1n}dc_{n-1,n}\int dc_{n-2,n},
\ee
since $(n-1)=0$ fixes $c_{n-2,n}=0$, hence the delta function $\de^2(\la_n-\la_Ic_{In})$ solves
\be
c_{n-1,n}=\frac{\<1n\>}{\<1,n-1\>},~c_{1n}=\frac{\<n-1,n\>}{\<n-1,1\>}. \labell{eq-13}
\ee
Putting every piece together,
\be
-\int d^3c_{In}\de^2(\la_n-\la_Ic_{In})\frac{(n-2)'(n-1)'}{(n-2)\und{(n-1)}(n)}
=\frac{1}{\<n-1,1\>}\frac{1}{c_{n-1,n}c_{1n}}=\frac{\<n-1,1\>}{\<n-1,n\>\<n1\>},
\ee
which matches the soft factor as promised.

Then what is going on in the rest parts of the Grassmannian integral?
Recall \eqref{eq-3}, note that delta functions
$\de^{2(n-4)}(\la_{\und{i}}-\la_Ic_{I\und{i}})$ are unaffected, while delta functions
$\de^{2\cdot3}(\tla_I+c_{I\und{i}}\tla_{\und{i}}+c_{In}\tla_n)$ turns into
\be
\bal
&\de^2(\tla_{n-1}+c_{n-1,\und{i}}\tla_{\und{i}}+c_{n-1,n}\tla_n)
\de^2(\tla_1+c_{1\und{i}}\tla_{\und{i}}+c_{1n}\tla_n)
\de^2(\tla_2+c_{2\und{i}}\tla_{\und{i}}+c_{2n}\tla_n)\\
=&\de^2\(\tla_{n-1}+\frac{\<1n\>}{\<1,n-1\>}\tla_n+c_{n-1,\und{i}}\tla_{\und{i}}\)
\de^2\(\tla_1+\frac{\<n-1,n\>}{\<n-1,1\>}\tla_n+c_{1\und{i}}\tla_{\und{i}}\)
\de^2(\tla_2+c_{2\und{i}}\tla_{\und{i}}), \labell{eq-17}
\eal
\ee
where the third one is also unaffected. Now let's recover the supersymmetric content, and we find
that it trivially imitates the expression above, namely
\be
\de^{4\cdot3}(\teta_I+c_{Ii}\teta_i)
=\de^4\(\teta_{n-1}+\frac{\<1n\>}{\<1,n-1\>}\teta_n+c_{n-1,\und{i}}\teta_{\und{i}}\)
\de^4\(\teta_1+\frac{\<n-1,n\>}{\<n-1,1\>}\teta_n+c_{1\und{i}}\teta_{\und{i}}\)
\de^4(\teta_2+c_{2\und{i}}\teta_{\und{i}}), \labell{eq-26}
\ee
Plug \eqref{eq-17} back into \eqref{eq-3} and also take \eqref{eq-26} into account, we just recover
the soft theorem for NMHV amplitudes in $\mathcal{N}=4$ SYM at tree level.
Explicitly, \eqref{eq-6} becomes
\be
\bal
&A^{[3]}_n\(\la_n\to\e\la_n\)\\
=&\frac{1}{\e^2}\frac{\<n-1,1\>}{\<n-1,n\>\<n1\>}
A^{[3]}_{n-1}\((\tla,\teta)_{n-1}\to(\tla,\teta)_{n-1}+\e\frac{\<1n\>}{\<1,n-1\>}(\tla,\teta)_n,\,
(\tla,\teta)_1\to(\tla,\teta)_1+\e\frac{\<n-1,n\>}{\<n-1,1\>}(\tla,\teta)_n\)\\
&+(\textrm{pure regular parts}), \labell{eq-4}
\eal
\ee
where $(\tla,\teta)_n$ is replaced by $\e(\tla,\teta)_n$ to manifest the soft divergence.
Here let's call the first term above `the singular parts',
but it in fact contains `mixed regular parts' after Taylor expansion in $\e$.
In contrast, the `pure regular parts' do not involve the $1/\e^2$ prefactor.
These pure regular parts correspond to terms
whose $A_L$'s are not 3-particle amplitudes in the BCFW construction, as the reader can look up
in \cite{Cachazo:14soft,Casali:14soft}. To see the singular parts directly, one can expand
the expression above as
\be
A^{[3]}_n\(\la_n\to\e\la_n\)=\(\frac{1}{\e^2}S^{(0)}+\frac{1}{\e}S^{(1)}\)A^{[3]}_{n-1}+O(\e^0),
\ee
where the leading soft factor and sub-leading soft operator are defined as
\be
S^{(0)}\equiv\frac{\<n-1,1\>}{\<n-1,n\>\<n1\>},~
S^{(1)}\equiv\frac{1}{\<n-1,n\>}\(\tla_n^{\dot{\alpha}}\frac{\partial}{\partial\tla_{n-1}^{\dot{\alpha}}}
+\teta_n^A\frac{\partial}{\partial\teta_{n-1}^A}\)
+\frac{1}{\<n1\>}\(\tla_n^{\dot{\alpha}}\frac{\partial}{\partial\tla_1^{\dot{\alpha}}}
+\teta_n^A\frac{\partial}{\partial\teta_1^A}\),
\ee
note that at order $\e^{-1}$ there are $\teta$ parts
in addition to the $\tla$ parts, which extends the sub-leading soft operator of
\eqref{eq-16} supersymmetrically\footnote{The same result given by the super BCFW construction
can be found in \cite{Liu:14soft}.}.
Still, this supersymmetric content behaves exactly like its bosonic counterpart.
Therefore we choose to suppress it in the following sections
for brevity. One can easily recover this content by imitating the $\tla$ part.

\section{More Extensions: N$^2$MHV and N$^3$MHV Amplitudes}
\labell{sec3}

Having accomplished the simplest case, now we would like to explore the
N$^2$MHV and N$^3$MHV amplitudes. From these two further examples,
the pattern for general N$^{k-2}$MHV amplitudes,
which will be revealed in section \ref{sec4}, starts to emerge.

To begin with the N$^2$MHV amplitude, recall
\eqref{eq-21} and \eqref{eq-22}, by applying the global residue
theorem, the relevant integral is
\be
\bal
\int\frac{g^{[4]}_n}{f'}\frac{1}{\und{F_7}\ldots\und{F_{n-1}}(\und{f_{n1}f_{n2}})}
&=-\int\frac{g^{[4]}_n}{\und{f'}}\frac{1}{\und{F_7}\ldots\und{F_{n-1}}(f_{n1}\und{f_{n2}})}\\
&=-\int\frac{g^{[4]}_n}{\und{(n-1)}(1)(3)}\frac{1}{\und{F_7}\ldots
\und{F_{n-1}}f_{n1}\und{(n-2)}(n)(\ldots)}-\ldots, \labell{24}
\eal
\ee
where $f'=(n-1)(1)(3)$ for conciseness, we also remind the reader
that $f_{n1}=(n-3)(\ldots)(\ldots)$ and $f_{n2}=(n-2)(n)(\ldots)$
where dots in parentheses denote the less important non-consecutive minors.
In the second line above, among $3^2$ choices of zero minors in $f'$
and $f_{n2}$, let's single out the term with $(n-1)=(n-2)=0$,
since other terms do not give singular contributions.

After the key integral is identified in \eqref{24},
following the similar recipe and using \eqref{eq-19}, we split the
integrand into the part of remaining $(n-1)$ particles and the part of
inverse soft factor, as treated in \eqref{eq-3}. Now let's focus on
\be
-\int d^4c_{In}\de^2(\la_n-\la_Ic_{In})
\frac{(n-3)'(n-2)'(n-1)'}{-(n-3)\und{(n-2)}\und{(n-1)}(n)}, \labell{eq-18}
\ee
as mentioned before, the reason to assign $(n-2)$ to be the second
zero minor is that this choice turns out to be the only singular
contribution in the soft limit. Its proof for general
$k$'s is given in appendix \ref{app2}. Note that there is another
minus sign appeared due to swapping the positions of $(n-1)$ and
$(n-2)$, since by default $(n-1)$ locates at the first place in
the sequence of zero minors.
The latter fact is also true for all $k$'s, as shown in appendix \ref{app1}.

To proceed, with the previous experience we choose the gauge
\be
C=\(\barr{ccccccccc}
\ldots & c_{n-2,n-3} & 1 & 0 & c_{n-2,n} & 0 & 0 & c_{n-2,3} & \ldots\\
\ldots & c_{n-1,n-3} & 0 & 1 & c_{n-1,n} & 0 & 0 & c_{n-1,3} & \ldots\\
\ldots & c_{1,n-3} & 0 & 0 & c_{1n} & 1 & 0 & c_{13} & \ldots\\
\ldots & c_{2,n-3} & 0 & 0 & c_{2n} & 0 & 1 & c_{23} & \ldots
\earr\),
\ee
where four columns $(n-2,n-1,1,2)$ have been fixed to be a unit matrix.
Hence the relevant minors are
\be
(n-3)'=-c_{2,n-3},~(n-2)'=1,~(n-1)'=-c_{n-2,3},
\ee
\be
(n-3)=\lv\barr{cc}
c_{1,n-3} & c_{1n}\\
c_{2,n-3} & c_{2n}
\earr\rv,~
(n-2)=-c_{2n},~(n-1)=-c_{n-2,n},~
(n)=\lv\barr{cc}
c_{n-2,n} & c_{n-2,3}\\
c_{n-1,n} & c_{n-1,3}
\earr\rv,
\ee
and the integration measure is
\be
\int d^4c_{In}=\int dc_{2n}dc_{1n}dc_{n-1,n}dc_{n-2,n}
=\int dc_{1n}dc_{n-1,n}\int dc_{2n}\int dc_{n-2,n}.
\ee
Pay attention to the order of $dc_{In}$'s as they in fact anticommute.
A little subtlety here is that we have to match the
orders of $dc_{In}$'s and zero minors, namely $dc_{2n}dc_{n-2,n}$
must be associated with $(n-2)(n-1)$, otherwise a sign factor will
arise due to altering the order of either $dc_{In}$'s or zero minors.
But nicely, there is no such a worry in this case and that's
the reason to adopt the reversed cyclic order for $dc_{In}$'s. Also note
that we always leave the integrations over $c_{1n}$ and $c_{n-1,n}$
to the last step, after performing the $(k-2)$ residue integrations.

The two integrations easily fix $c_{2n}=c_{n-2,n}=0$, and using
the remaining two delta functions in \eqref{eq-18} we find the solution \eqref{eq-13}.
Hence the final result is
\be
\frac{1}{\<n-1,1\>}\frac{1}{c_{n-1,n}c_{1n}}=\frac{\<n-1,1\>}{\<n-1,n\>\<n1\>},
\ee
and the anti-holomorphic spinor pair $(\tla_{n-1},\tla_1)$ is
deformed identically as \eqref{eq-17}, regardless
of the increase of $k$. Not surprisingly, the soft theorem is again
recovered for the N$^2$MHV case.

Next we move on to the case of N$^3$MHV amplitudes, because it contains
a non-trivial feature that cannot be seen in the former case. By applying the global residue theorem,
the relevant integral is\footnote{This form for $k=5$ will be explained in appendix \ref{app1}.}
\be
\int\frac{g^{[5]}_n}{f'}\frac{1}{\und{F_8}\ldots\und{F_{n-1}}(\und{f_{n1}f_{n2}f_{n3}})}
=-\int\frac{g^{[5]}_n}{\und{f'}}\frac{1}{\und{F_8}\ldots\und{F_{n-1}}(f_{n1}\und{f_{n2}f_{n3}})},
\ee
where $f'=(n-1)(1)(3)$. In the RHS above, among $3^3$ choices
of zero minors in $f'$, $f_{n2}$ and $f_{n3}$, as you may guess, we single out
the one picking $(n-1)$, $(n-2)$ and $(n-3)$ respectively,
where following expressions of $f_{nj}$'s are used:
\be
f_{n1}=(n-4)(\ldots)(\ldots),~f_{n2}=(n-3)(\ldots)(\ldots),~f_{n3}=(n-2)(n)(\ldots).
\ee
Following the pattern of \eqref{eq-18} and using \eqref{eq-19}, let's calculate
\be
-\int d^5c_{In}\de^2(\la_n-\la_Ic_{In})
\frac{(n-4)'(n-3)'(n-2)'(n-1)'}{(-)^2(n-4)\und{(n-3)}\und{(n-2)}\und{(n-1)}(n)}, \labell{eq-5}
\ee
the reason to assign $(n-3)$ to be the third zero minor is the same
as previous. Here, a sign factor also arises when $(n-1)$ is pulled
through $(n-3)(n-2)$, since only the zero minors care about their
order while others trivially commute.

To proceed as trickily as before, we choose the gauge
\be
C=\(\barr{ccccccccccc}
\ldots & c_{n-2,n-4} & c_{n-2,n-3} & 1 & 0 & c_{n-2,n} & 0 & 0 & 0 & c_{n-2,4} & \ldots\\
\ldots & c_{n-1,n-4} & c_{n-1,n-3} & 0 & 1 & c_{n-1,n} & 0 & 0 & 0 & c_{n-1,4} & \ldots\\
\ldots & c_{1,n-4} & c_{1,n-3} & 0 & 0 & c_{1n} & 1 & 0 & 0 & c_{14} & \ldots\\
\ldots & c_{2,n-4} & c_{2,n-3} & 0 & 0 & c_{2n} & 0 & 1 & 0 & c_{24} & \ldots\\
\ldots & c_{3,n-4} & c_{3,n-3} & 0 & 0 & c_{3n} & 0 & 0 & 1 & c_{34} & \ldots
\earr\),
\ee
where five columns $(n-2,n-1,1,2,3)$ have been fixed to be a unit matrix.
Compare this choice with those in NMHV and N$^2$MHV cases,
keen eyes will immediately see the pattern: We always fix $k$
columns $(n-2,n-1,1,2,\ldots,k-2)$ to be a unit matrix. Hence the relevant minors are
\be
(n-4)'=\lv\barr{cc}
c_{2,n-4} & c_{2,n-3}\\
c_{3,n-4} & c_{3,n-3}
\earr\rv,~
(n-3)'=c_{3,n-3},~(n-2)'=1,~(n-1)'=c_{n-2,4},
\ee
\be
\bal
&(n-4)=\lv\barr{ccc}
c_{1,n-4} & c_{1,n-3} & c_{1n}\\
c_{2,n-4} & c_{2,n-3} & c_{2n}\\
c_{3,n-4} & c_{3,n-3} & c_{3n}
\earr\rv,~
(n-3)=\lv\barr{cc}
c_{2,n-3} & c_{2n}\\
c_{3,n-3} & c_{3n}
\earr\rv,\\
&(n-2)=c_{3n},~(n-1)=-c_{n-2,n},~
(n)=-\lv\barr{cc}
c_{n-2,n} & c_{n-2,4}\\
c_{n-1,n} & c_{n-1,4}
\earr\rv.
\eal
\ee
and the integration measure is
\be
\int d^5c_{In}=\int dc_{3n}dc_{2n}dc_{1n}dc_{n-1,n}dc_{n-2,n}=-\int dc_{1n}dc_{n-1,n}
\int dc_{2n}dc_{3n}\int dc_{n-2,n},
\ee
where the order above is chosen to fit $(n-3)(n-2)(n-1)$. We must
start the integrations from the rightmost, so the residue at
$(n-3)=0$ must be evaluated after finishing $(n-2)$ and $(n-1)$. In
this way the cancelation of all other un-localized $c_{I\und{i}}$'s
is guaranteed, as the relevant minors factorize with particular zero
entries. This is a general pattern of N$^{k-2}$MHV amplitudes, but it
only starts to emerge from the N$^3$MHV case.

As expected, the final result of \eqref{eq-5} is
\be
\frac{1}{\<n-1,1\>}\frac{1}{c_{n-1,n}c_{1n}}=\frac{\<n-1,1\>}{\<n-1,n\>\<n1\>},
\ee
with $c_{2n}=c_{3n}=c_{n-2,n}=0$,
spinor pair $(\tla_{n-1},\tla_1)$ is deformed identically as \eqref{eq-17}.
Once more, the soft theorem is recovered for the N$^3$MHV case.

\section{General N$^{k-2}$MHV Amplitudes}
\labell{sec4}

In the previous case, we actually presume that N$^{k-2}$MHV amplitudes have a universal structure,
where $(n-1)$ and $(n)$ play special roles, namely
\be
A^{[k]}_n=\int\frac{g^{[k]}_n}{f'}\frac{1}{F_{k+3}\ldots F_n},~f'=(n-1)(1)(3),~
F_i=f_{i1}\ldots f_{i,k-2}, \labell{eq-10}
\ee
which is constructed by conjugation in appendix \ref{app1}. Applying the global residue theorem, yields
\be
\int\frac{g^{[k]}_n}{f'}\frac{1}{\und{F_{k+3}}\ldots\und{F_{n-1}}(\und{f_{n1}f_{n2}\ldots f_{n,k-2}})}
=-\int\frac{g^{[k]}_n}{\und{f'}}
\frac{1}{\und{F_{k+3}}\ldots\und{F_{n-1}}(f_{n1}\und{f_{n2}}\ldots\und{f_{n,k-2}})}, \labell{eq-12}
\ee
where each underlined $F_i$ enforces all $(k-2)$ $f_{ij}$'s it
contains to be zero, and each $f_{ij}$ contributes one zero minor at a time
respectively. The form of $f_{nj}$'s is given by
\be
f_{n1}=(n-k+1)(\ldots)(\ldots),~\ldots~
f_{n,k-3}=(n-3)(\ldots)(\ldots),~f_{n,k-2}=(n-2)(n)(\ldots).
\ee
Among $3^{k-2}$ choices of zero minors in the RHS of \eqref{eq-12},
we single out the one picking
$(n-1)=(n-k+2)=(n-k+3)=\ldots=(n-2)=0$, selected from $f'$ and
$f_{nj}$'s with $j=2,\ldots,k-2$. This one is the only term that has singular
contribution in the holomorphic soft limit.

To proceed, we choose the gauge
\be
C=\(\barr{cccccccccccccccccc}
\ldots & c_{n-2,n-k+1} & c_{n-2,n-k+2} & \ldots & c_{n-2,n-4} & c_{n-2,n-3} & 1 & 0 & c_{n-2,n} & 0 & 0
& \ldots & 0 & 0 & c_{n-2,k-1} & \ldots\\
\ldots & c_{n-1,n-k+1} & c_{n-1,n-k+2} & \ldots & c_{n-1,n-4} & c_{n-1,n-3} & 0 & 1 & c_{n-1,n} & 0 & 0
& \ldots & 0 & 0 & c_{n-1,k-1} & \ldots\\
\ldots & c_{1,n-k+1} & c_{1,n-k+2} & \ldots & c_{1,n-4} & c_{1,n-3} & 0 & 0 & c_{1n} & 1 & 0
& \ldots & 0 & 0 & c_{1,k-1} & \ldots\\
\ldots & c_{2,n-k+1} & c_{2,n-k+2} & \ldots & c_{2,n-4} & c_{2,n-3} & 0 & 0 & c_{2n} & 0 & 1
& \ldots & 0 & 0 & c_{2,k-1} & \ldots\\
\vdots & \vdots & \vdots & \ddots & \vdots & \vdots & \vdots & \vdots & \vdots & \vdots & \vdots & \ddots
& \vdots & \vdots & \vdots & \vdots\\
\ldots & c_{k-3,n-k+1} & c_{k-3,n-k+2} & \ldots & c_{k-3,n-4} & c_{k-3,n-3} & 0 & 0 & c_{k-3,n} & 0 & 0
& \ldots & 1 & 0 & c_{k-3,k-1} & \ldots\\
\ldots & c_{k-2,n-k+1} & c_{k-2,n-k+2} & \ldots & c_{k-2,n-4} & c_{k-2,n-3} & 0 & 0 & c_{k-2,n} & 0 & 0
& \ldots & 0 & 1 & c_{k-2,k-1} & \ldots
\earr\),
\ee
where $k$ columns $(n-2,n-1,1,2,\ldots,k-3,k-2)$ have been fixed to be a unit matrix.
Next let's calculate the following integral
\be
-\int d^kc_{In}\de^2(\la_n-\la_Ic_{In})\frac{(n-k+1)'(n-k+2)'(n-k+3)'\ldots(n-1)'}
{(-)^{k-3}(n-k+1)\und{(n-k+2)}\und{(n-k+3)}\ldots\und{(n-2)}\und{(n-1)}(n)}, \labell{eq-9}
\ee
which is obtained by extending \eqref{eq-5} for a generic $k$ after using \eqref{eq-19}.
The sign factor in the denominator arises when
$(n-1)$ is pulled through other $(k-3)$ zero minors. As previous,
the only singular contribution is from the sequence of zero minors selected above.

In the chosen gauge, the relevant minors are
\be
\bal
(n-k+1)'&=(-)^{(k-3)\cdot3}\lv\barr{ccc}
c_{2,n-k+1} & \ldots & c_{2,n-3}\\
\vdots & \ddots & \vdots\\
c_{k-2,n-k+1} & \ldots & c_{k-2,n-3}
\earr\rv,\\
(n-k+2)'&=(-)^{(k-4)\cdot4}\lv\barr{ccc}
c_{3,n-k+2} & \ldots & c_{3,n-3}\\
\vdots & \ddots & \vdots\\
c_{k-2,n-k+2} & \ldots & c_{k-2,n-3}
\earr\rv,\\
&~\vdots\\
(n-4)'&=(-)^{2(k-2)}\lv\barr{cc}
c_{k-3,n-4} & c_{k-3,n-3}\\
c_{k-2,n-4} & c_{k-2,n-3}
\earr\rv,\\
(n-3)'&=(-)^{1(k-1)}c_{k-2,n-3},~
(n-2)'=1,~
(n-1)'=(-)^{k-1}c_{n-2,k-1}, \labell{eq-8}
\eal
\ee
as well as
\be
\bal
(n-k+1)&=(-)^{(k-2)\cdot2}\lv\barr{cccc}
c_{1,n-k+1} & \ldots & c_{1,n-3} & c_{1n}\\
\vdots & \ddots & \vdots & \vdots\\
c_{k-2,n-k+1} & \ldots & c_{k-2,n-3} & c_{k-2,n}
\earr\rv,\\
(n-k+2)&=(-)^{(k-3)\cdot3}\lv\barr{cccc}
c_{2,n-k+2} & \ldots & c_{2,n-3} & c_{2n}\\
\vdots & \ddots & \vdots & \vdots\\
c_{k-2,n-k+2} & \ldots & c_{k-2,n-3} & c_{k-2,n}
\earr\rv,\\
&~\vdots\\
(n-4)&=(-)^{3(k-3)}\lv\barr{ccc}
c_{k-4,n-4} & c_{k-4,n-3} & c_{k-4,n}\\
c_{k-3,n-4} & c_{k-3,n-3} & c_{k-3,n}\\
c_{k-2,n-4} & c_{k-2,n-3} & c_{k-2,n}
\earr\rv,\\
(n-3)&=(-)^{2(k-2)}\lv\barr{cc}
c_{k-3,n-3} & c_{k-3,n}\\
c_{k-2,n-3} & c_{k-2,n}
\earr\rv,\\
(n-2)&=(-)^{1(k-1)}c_{k-2,n},~
(n-1)=-c_{n-2,n},\\
(n)&=(-)^{k-2}\lv\barr{cc}
c_{n-2,n} & c_{n-2,k-1}\\
c_{n-1,n} & c_{n-1,k-1}
\earr\rv. \labell{eq-7}
\eal
\ee
And the integration measure is
\be
\bal
\int d^kc_{In}&=\int dc_{k-2,n}dc_{k-3,n}\ldots dc_{2n}dc_{1n}dc_{n-1,n}dc_{n-2,n}\\
&=(-)^{1+2+\ldots+(k-4)}\int dc_{1n}dc_{n-1,n}\int dc_{2n}\ldots dc_{k-3,n}dc_{k-2,n}\int dc_{n-2,n},
\eal
\ee
note that we have reversed the order of $dc_{k-2,n}dc_{k-3,n}\ldots dc_{2n}$ to fit
$(n-k+2)(n-k+3)\ldots(n-2)$, hence a sign factor arises. The $(k-2)$ residue integrations at
$(n-k+2)=(n-k+3)=\ldots=(n-2)=(n-1)=0$
fix $(k-2)$ $c_{In}$'s to be zero when proceeded from right to left,
while $c_{n-1,n}$ and $c_{1n}$ are localized by the delta function
$\de^2(\la_n-\la_Ic_{In})$ in the last step.

After that, \eqref{eq-7} reduces to
\be
\bal
(n-k+1)&=(-)^{(k-2)\cdot2}\lv\barr{cccc}
c_{1,n-k+1} & \ldots & c_{1,n-3} & c_{1n}\\
c_{2,n-k+1} & \ldots & c_{2,n-3} & \und{c_{2n}}\\
\vdots & \ddots & \vdots & \vdots\\
c_{k-2,n-k+1} & \ldots & c_{k-2,n-3} & \und{c_{k-2,n}}
\earr\rv
=(-)^{(k-2)\cdot2+(k-3)}c_{1n}\lv\barr{ccc}
c_{2,n-k+1} & \ldots & c_{2,n-3}\\
\vdots & \ddots & \vdots\\
c_{k-2,n-k+1} & \ldots & c_{k-2,n-3}
\earr\rv,\\
(n-k+2)&=(-)^{(k-3)\cdot3}\lv\barr{cccc}
c_{2,n-k+2} & \ldots & c_{2,n-3} & c_{2n}\\
c_{3,n-k+2} & \ldots & c_{3,n-3} & \und{c_{3n}}\\
\vdots & \ddots & \vdots & \vdots\\
c_{k-2,n-k+2} & \ldots & c_{k-2,n-3} & \und{c_{k-2,n}}
\earr\rv
=(-)^{(k-3)\cdot3+(k-4)}c_{2n}\lv\barr{cccc}
c_{3,n-k+2} & \ldots & c_{3,n-3}\\
\vdots & \ddots & \vdots\\
c_{k-2,n-k+2} & \ldots & c_{k-2,n-3}
\earr\rv,\\
&~\vdots\\
(n-4)&=(-)^{3(k-3)}\lv\barr{ccc}
c_{k-4,n-4} & c_{k-4,n-3} & c_{k-4,n}\\
c_{k-3,n-4} & c_{k-3,n-3} & \und{c_{k-3,n}}\\
c_{k-2,n-4} & c_{k-2,n-3} & \und{c_{k-2,n}}
\earr\rv
=(-)^{3(k-3)+2}c_{k-4,n}\lv\barr{cc}
c_{k-3,n-4} & c_{k-3,n-3}\\
c_{k-2,n-4} & c_{k-2,n-3}
\earr\rv,\\
(n-3)&=(-)^{2(k-2)}\lv\barr{cc}
c_{k-3,n-3} & c_{k-3,n}\\
c_{k-2,n-3} & \und{c_{k-2,n}}
\earr\rv
=(-)^{2(k-2)+1}c_{k-3,n}\,c_{k-2,n-3},\\
(n-2)&=(-)^{1(k-1)}c_{k-2,n},~
(n-1)=-c_{n-2,n},\\
(n)&=(-)^{k-2}\lv\barr{cc}
\und{c_{n-2,n}} & c_{n-2,k-1}\\
c_{n-1,n} & c_{n-1,k-1}
\earr\rv
=(-)^{k-1}c_{n-1,n}\,c_{n-2,k-1}. \labell{eq-23}
\eal
\ee
Now the pattern is clear:
$(n-1)=0$ fixes $c_{n-2,n}=0$, hence $(n)$ factorizes into
$(-)^{k-1}c_{n-1,n}\,c_{n-2,k-1}$. $(n-2)=0$ fixes $c_{k-2,n}=0$,
hence $(n-3)$ factorizes into $(-)^{2(k-2)+1}c_{k-3,n}\,c_{k-2,n-3}$,
and a further integration fixes $c_{k-3,n}=0$.
Having $c_{k-2,n}=c_{k-3,n}=0$, $(n-4)$ again factorizes.
This pattern will repeat to the `top' minor $(n-k+1)$.
Therefore the correct order to proceed residue
integrations is crucial. Combine the result above with \eqref{eq-8}
and plug them back into \eqref{eq-9}, all un-localized $c_{I\und{i}}$'s
cancel, and we reach the longing answer:
\be
\frac{\<n-1,1\>}{\<n-1,n\>\<n1\>}\times\textrm{Sign},
\ee
where
\be
\textrm{Sign}=\frac{(-)(-)^{1+2+\ldots+(k-4)}(-)^{k-1}(-)^{k-1}\times(-)^{2(k-2)}\ldots(-)^{(k-3)\cdot3}}
{(-)^{k-3}(-)(-)^{k-1}(-)^{k-1}\times(-)^{2(k-2)+1}\ldots(-)^{(k-3)\cdot3+(k-4)}(-)^{(k-2)\cdot2+(k-3)}}=1,
\ee
which is not a coincidence, but a consequence of cautiously chosen conventions.
Since $c_{2n}=c_{3n}=\ldots=c_{k-2,n}=c_{n-2,n}=0$,
spinor pair $(\tla_{n-1}, \tla_1)$ is deformed identically as \eqref{eq-17}.
The soft theorem for general N$^{k-2}$MHV amplitudes is now proved.

\section{Conclusion}
\labell{conclu}

In this note, we see that although it is already difficult to get explicit expressions for N$^2$MHV
amplitudes by performing all residue integrations, let alone
general N$^{k-2}$MHV amplitudes, still in the soft limit, with sufficient
tricks one is able to find the soft theorem for all $k$'s while
keeping the Grassmannian contour integrations of $A^{[k]}_{n-1}$ unsolved.
In proving this relation, a judicious choice of the Grassmannian gauge leads to
considerable simplification. Besides, the irrelevant role of non-consecutive minors,
which deserves to be emphasized, greatly reduces the complexity of the integrand structure.

\section*{Acknowledgement}

The author is grateful to Prof. Bo Feng for his ardent guidance
during the project. This work is supported by Chinese NSF funding
under contracts No.11031005, No.11135006 and No.11125523.

\appendix
\section{Conjugation Construction of General N$^{k-2}$MHV Amplitudes}
\labell{app1}

Given an amplitude $A^{[k]}_n$, one can construct two other amplitudes:
$A^{[k]}_{n+m}$ with $m$ extra particles of positive helicity,
and $A^{[k+l]}_n$ with $l$ positive helicities flipped.
It is easy to construct $A^{[k]}_{n+m}$ by applying the inverse soft
operation (`add one particle at a time') successively.
However, for $A^{[k+l]}_n$ it is not so straightforward
and we need to use a trick of conjugation.

Assume that $k=3$ and $n=6+l$, the conjugation gives
\be
A^{[3+l]}_{6+l}=\ov{A^{[3]}_{6+l}},
\ee
then the general $A^{[3+l]}_n$ can be obtained from $A^{[3+l]}_{6+l}$, which we named as the `seed amplitude',
since the inverse soft operation will grow it into amplitudes for all $n$'s while fixing $k'=3+l$.

Based on this observation, we now present an
incomplete approach to construct amplitudes for all
$k$'s. It is incomplete because we will leave the unnecessary part,
which involves overwhelming products of non-consecutive minors, unspecified in the derivation.
The complete formula can be found in \cite{BTVW:10trees}.

Let's start by rewriting \eqref{eq-10} as
\be
A^{[k]}_n=\int\frac{g^{[k]}_n}{[(n-1)(1)(3)]_k}\frac{1}{[F_{k+3}\ldots F_n]_k},~
[F_i]_k=[f_{i1}\ldots f_{i,k-2}]_k, \labell{eq-11}
\ee
where $[\ldots]_k$ is a collective type label, for instance, $[abc]_k=a_kb_kc_k$,
and it is introduced to distinguish `the same functions' of different $k$'s.
Setting $k=3$ and $n=6+l$, then the seed amplitude is
\be
A^{[3+l]}_{6+l}=\int\frac{\ov{g^{[3]}_{6+l}}}{[(8+l)(4)(6)]_{3+l}}
\frac{1}{\[\,\ov{f_{6}}\ldots\ov{f_{6+l}}\,\]_{3+l}},
\ee
after recalling that the conjugate of $(l)_k=(l\,l+1\,\ldots\,l+k-1)$ is
$(l+k)_{n-k}=(l+k\,l+k+1\,\ldots\,l-1)$. To return to the analogous
form of \eqref{eq-11}, we use the cyclic invariance to perform the
following shift of labels: $i\to i-3$, now
\be
A^{[3+l]}_{6+l}=\int\frac{\s\ov{g^{[3]}_{6+l}}}{[(5+l)(1)(3)]_{3+l}}
\frac{1}{\[\s\ov{f_{6}}\ldots\s\ov{f_{6+l}}\,\]_{3+l}},
\ee
where the $\hat{s}$ operator denotes the cyclic shift.
Define new variables via $\s\ov{g^{[3]}_{6+l}}\equiv g^{[3+l]}_{6+l}$, and
\be
\s\ov{f_{6}}\ldots\s\ov{f_{6+l}}\equiv f_{{6+l},1}\ldots f_{{6+l},1+l}=F_{6+l}, \labell{eq-25}
\ee
hence we achieve
\be
A^{[3+l]}_{6+l}=\int\frac{g^{[3+l]}_{6+l}}{[(5+l)(1)(3)]_{3+l}}\frac{1}{[F_{6+l}]_{3+l}},
\ee
extending from $(6+l)$ to generic $n$ by using `add one particle a at time' successively, yields
\be
A^{[3+l]}_n=\int\frac{g^{[3+l]}_n}{[(n-1)(1)(3)]_{3+l}}\frac{1}{[F_{6+l}\ldots F_n]_{3+l}},~
[F_i]_{3+l}=[f_{i1}\ldots f_{i,1+l}]_{3+l},
\ee
which returns to \eqref{eq-11}. From the initial amplitude $A^{[3]}_6$, we know that each $f_{ij}$
contains three minors and this is true for all $n$'s and $k$'s, as implied by this construction.
It is also confirmed by the first non-trivial example of N$^2$MHV amplitudes,
as mentioned in section \ref{intro}.

We have left the explicit expressions of $f_{ij}$'s unspecified,
but for the sake of proving the soft theorem,
one key fact must be clarified: The form of $f_{nj}$'s is given by
\be
f_{n1}=[(n-k+1)(\ldots)(\ldots)]_k,~\ldots~
f_{n,k-3}=[(n-3)(\ldots)(\ldots)]_k,~f_{n,k-2}=[(n-2)(n)(\ldots)]_k. \labell{eq-24}
\ee
Obviously the minor $(n)$ plays a special role above, in addition to the special minor $(n-1)$.
Also note that all consecutive minors involving column $n$, except $(n-1)$, must locate in
$F_n=f_{n1}\ldots f_{n,k-2}$.
For given generic $n$ and $k$, the proof of this arrangement is the following.

By default, this arrangement trivially extends to the $(n+m)$ case while fixing $k$,
then let's fix $n=6+l$ and replace $k=3$ by $k'=3+l$.
Firstly \eqref{eq-24} is valid for all $n$'s in cases of $k=3$ and $k=4$,
as confirmed in section \ref{intro}. To extend it by induction, note that
in process \eqref{eq-25} of constructing $F_{6+l}$ from $f_6\ldots f_{6+l}$,
the operation of conjugation followed by label shift exactly preserves the minor labels,
while $k=3$ is replaced by $k'=3+l$.
Explicitly, we find the following form identical to \eqref{eq-24}, namely
\be
f_{n,1}=[(n-k'+1)(\ldots)(\ldots)]_{k'},~\ldots~
f_{n,k'-3}=[(n-3)(\ldots)(\ldots)]_{k'},~f_{n,k'-2}=[(n-2)(n)(\ldots)]_{k'},
\ee
where $n$ and $k'$ are kept instead of $l$. Therefore the proof is done.

The order of minors involving column $n$ in \eqref{eq-24} is
$(n-k+1)(n-k+2)\ldots(n-2)(n)$, which justifies the order in \eqref{eq-9}.

\section{Pure Regular Parts}
\labell{app2}

In this part, let's show that after using the global residue
theorem, only one term has singular contribution. For the reader's
convenience, we write \eqref{eq-12} again here
\be
\int\frac{g^{[k]}_n}{f'}\frac{1}{\und{F_{k+3}}\ldots\und{F_{n-1}}(\und{f_{n1}f_{n2}\ldots f_{n,k-2}})}
=-\int\frac{g^{[k]}_n}{\und{f'}}
\frac{1}{\und{F_{k+3}}\ldots\und{F_{n-1}}(f_{n1}\und{f_{n2}}\ldots\und{f_{n,k-2}})},
\ee
with $f'=(n-1)(1)(3)$ and
\be
f_{n1}=(n-k+1)(\ldots)(\ldots),~\ldots~
f_{n,k-3}=(n-3)(\ldots)(\ldots),~f_{n,k-2}=(n-2)(n)(\ldots).
\ee
It has been claimed that the only singular contribution in the soft limit is from the particular sequence
of zero minors selected above, namely $(n-1)(n-k+2)(n-k+3)\ldots(n-2)$,
while all other choices give regular terms.

To see why, let's recall the origin of soft divergence in the Grassmannian formulation.
From section \ref{sec4}, we know that the $(k-2)$ residue integrations
enforce $c_{2n}=c_{3n}=\ldots=c_{k-2,n}=c_{n-2,n}=0$,
leaving only $c_{n-1,n}$ and $c_{1n}$ non-vanishing. Write \eqref{eq-13} again here,
\be
c_{n-1,n}=\frac{\<1n\>}{\<1,n-1\>},~c_{1n}=\frac{\<n-1,n\>}{\<n-1,1\>},
\ee
or equivalently,
\be
\la_n=\la_{n-1}c_{n-1,n}+\la_1c_{1n}.
\ee
It is natural to conceive that if there are some extra pieces besides $\la_n$ on the LHS of
this equation, $c_{n-1,n}$ and $c_{1n}$ would be `protected' from vanishing in the limit
$\la_n\to\e\la_n$, and hence the denominator involving $c_{n-1,n}$ and $c_{1n}$ would not cause
divergence since it is non-zero.

Favorably this is the right hint to catch. Rewrite the $\la_n$ equation
before localizing all $c_{In}$'s as
\be
\la_n-\la_{\und{I}}c_{\und{I}n}=\la_{n-1}c_{n-1,n}+\la_1c_{1n},
\ee
where $\und{I}=n-2,2,3,\ldots,k-2$. Any selected sequence of zero minors other than
$(n-k+2)(n-k+3)\ldots(n-2)(n-1)$ fails to localize all $(k-2)$ $c_{\und{I}n}$'s to be zero,
since these are the only minors involving column $n$ besides $(n-k+1)$ and $(n)$,
while the latter two are localized by the delta function $\de^2(\la_n-\la_Ic_{In})$ as always.
Consider the extreme case where all $(k-2)$ $c_{\und{I}n}$'s are not localized by these $(k-2)$ zero minors,
then they must be localized by other constraints in the full
Grassmannian integral. The relevant part in the integral involving $c_{In}$'s is
\be
\bal
&\int d^kc_{In}\de^2(\la_n-\la_Ic_{In})
\frac{\de^{2\cdot k}(\tla_I+c_{I\und{i}}\tla_{\und{i}}+c_{In}\tla_n)}
{(n-k+1)(n-k+2)(n-k+3)\ldots(n-2)(n-1)(n)}\\
=&\frac{1}{\<n-1,1\>}\int d^{k-2}c_{\und{I}n}
\frac{\de^{2\cdot(k-2)}(\tla_{\und{I}}+c_{\und{I}\und{i}}\tla_{\und{i}}+c_{\und{I}n}\tla_n)
\de^2(\tla_{n-1}+c_{n-1,\und{i}}\tla_{\und{i}}+c_{n-1,n}\tla_n)
\de^2(\tla_1+c_{1\und{i}}\tla_{\und{i}}+c_{1n}\tla_n)}
{(n-k+1)(n-k+2)(n-k+3)\ldots(n-2)(n-1)(n)}, \labell{eq-20}
\eal
\ee
where
\be
c_{n-1,n}=\frac{\<1n\>-\<1\und{I}\>c_{\und{I}n}}{\<1,n-1\>},~
c_{1n}=\frac{\<n-1,n\>-\<n-1,\und{I}\>c_{\und{I}n}}{\<n-1,1\>}.
\ee
Since $2(k-2)\geq k-2$, all $c_{\und{I}n}$'s can be localized by delta functions selected
from the $(k-2)$ constraints on their associated spinors.
These values of $c_{\und{I}n}$'s only depend on
anti-holomorphic spinors, hence they are free from the holomorphic soft limit.
When $c_{\und{I}n}$'s are finite, $c_{n-1,n}$ and $c_{1n}$ also remain finite
even though $\la_n\to\e\la_n$, hence the possible divergence
caused by these two variables safely dissolves in \eqref{eq-20}.

When $(k-2)$ $c_{\und{I}n}$'s are partially localized, the argument above works analogously.
This is why only one selected sequence of $(k-2)$ zero minors can cause soft divergence, as this choice delicately
removes all `protections' against pushing $c_{n-1,n}$ and $c_{1n}$ to zero.



\begin{thebibliography}{11111111}


\bibitem{Low:54}
  F. E. Low, Phys. Rev. {\bf 96}, 1428 (1954);\\
  M. Gell-Mann and M. L. Goldberger, Phys. Rev. {\bf 96}, 1433 (1954);\\
  S. Saito, Phys. Rev. {\bf 184}, 1894 (1969).


\bibitem{Low:58}
 F. E. Low, Phys. Rev. {\bf 110}, 974 (1958).


\bibitem{Weinberg}
 S. Weinberg, Phys. Rev. {\bf 135} (1964) B1049;\\
 S. Weinberg, Phys. Rev. {\bf 140} (1965) B516.




\bibitem{Cachazo:14soft}
  F. Cachazo and A. Strominger,
  ``Evidence for a New Soft Graviton Theorem,''
  arXiv:1404.4091 [hep-th].


\bibitem{Casali:14soft}
  E. Casali,
  JHEP {\bf 1408}, 077 (2014)
  [arXiv:1404.5551 [hep-th]].


\bibitem{Arkani-Hamed:09duality}
  N. Arkani-Hamed, F. Cachazo, C. Cheung and J. Kaplan,
  JHEP {\bf 1003}, 020 (2010)
  [arXiv:0907.5418 [hep-th]].


\bibitem{Arkani-Hamed:09unification}
  N. Arkani-Hamed, J. Bourjaily, F. Cachazo and J. Trnka,
  JHEP {\bf 1101}, 049 (2011)
  [arXiv:0912.4912 [hep-th]].


\bibitem{NVW:09etude}
  D. Nandan, A. Volovich and C. Wen,
  JHEP {\bf 1007}, 061 (2010)
  [arXiv:0912.3705 [hep-th]].


\bibitem{BTVW:10trees}
  J. Bourjaily, J. Trnka, A. Volovich and C. Wen,
  JHEP {\bf 1101}, 038 (2011)
  [arXiv:1006.1899 [hep-th]].


\bibitem{Liu:14soft}
  Z. -W. Liu,
  ``Soft theorems in maximally supersymmetric theories,''
  arXiv:1410.1616 [hep-th].




\bibitem{BLPR:14constrain}
  J. Broedel, M. de Leeuw, J. Plefka and M. Rosso,
  Phys. Rev. D {\bf 90} 065024 (2014)
  [arXiv:1406.6574 [hep-th]].


\bibitem{BDVN:14lowenergy}
  Z. Bern, S. Davies, P. Di Vecchia and J. Nohle,
  ``Low-Energy Behavior of Gluons and Gravitons from Gauge Invariance,''
  arXiv:1406.6987 [hep-th].


\bibitem{White:14diagram}
  C. D. White,
  Phys. Rev. B {\bf 737} 216222 (2014)
  [arXiv:1406.7184 [hep-th]].


\bibitem{Larkoski:14conformal}
  A. J. Larkoski,
  Phys. Rev. D {\bf 90} 087701 (2014)
  [arXiv:1405.2346 [hep-th]].


\bibitem{SV:14arbitrary}
  B. U. W. Schwab and A. Volovich,
  Phys. Rev. Lett. {\bf 113} 101601 (2014)
  [arXiv:1404.7749 [hep-th]].


\bibitem{Afkhami-Jeddi:14arbitrary}
  N. Afkhami-Jeddi,
  ``Soft Graviton Theorem in Arbitrary Dimensions,''
  arXiv:1405.3533 [hep-th].


\bibitem{Zlotnikov:14arbitrary}
  M. Zlotnikov,
  ``Sub-sub-leading soft-graviton theorem in arbitrary dimension,''
  arXiv:1407.5936 [hep-th].


\bibitem{KR:14arbitrary}
  C. Kalousios and F. Rojas,
  ``Next to subleading soft-graviton theorem in arbitrary dimensions,''
  arXiv:1407.5982 [hep-th].


\bibitem{BDN:14oneloop}
  Z. Bern, S. Davies and J. Nohle,
  ``On Loop Corrections to Subleading Soft Behavior of Gluons and Gravitons,''
  arXiv:1405.1015 [hep-th].


\bibitem{HHW:14loop}
  S. He, Y. -t. Huang and C. Wen,
  ``Loop Corrections to Soft Theorems in Gauge Theories and Gravity,''
  arXiv:1405.1410 [hep-th].


\bibitem{Cachazo:14renormalized}
  F. Cachazo and E. Y. Yuan,
  ``Are Soft Theorems Renormalized,''
  arXiv:1405.3413 [hep-th].


\bibitem{Schwab:14string}
  B. U. W. Schwab,
  JHEP {\bf 1408}, 062 (2011)
  [arXiv:1406.4172 [hep-th]].


\bibitem{BHHW:14more}
  M. Bianchi, S. He, Y. -t. Huang and C. Wen,
  ``More on Soft Theorems: Trees, Loops and Strings,''
  arXiv:1406.5155 [hep-th].


\bibitem{ACS:14null}
  T. Adamo, E. Casali and D. Skinner,
  ``Perturbative gravity at null infinity,''
  arXiv:1405.5122 [hep-th].


\bibitem{GLM:14null}
  Y. Geyer, A. E. Lipstein and L. Mason,
  ``Ambitwistor strings at null infinity and subleading soft limits,''
  arXiv:1406.1462 [hep-th].


\bibitem{DFFW:14KLT}
  Y. -J. Du, B. Feng, C. -H. Fu, Y. Wang,
  ``Note on Soft Graviton theorem by KLT Relation,''
  arXiv:1408.4179 [hep-th].


\end{thebibliography}
\end{document}